\documentclass[useAMS,usenatbib,referee]{mn2e}
\usepackage{bm}
\input epsf.tex
\usepackage{graphicx}
\def\b{\bm}

\renewcommand{\b}[1]{\mbox{\boldmath $#1$}}

\def\bqy{\begin{eqnarray}}
\def\eqy{\end{eqnarray}}
\def\bqyn{\begin{eqnarray*}}
\def\eqyn{\end{eqnarray*}}
\def\bq{\begin{equation}}
\def\eq{\end{equation}}
\def\oline{\overline}

\title{Magnetic Transport On The Solar Atmosphere By Turbulent Ambipolar Diffusion}

\author[V. Krishan and S. Masuda]{V. Krishan$^{1,2}$
and S. Masuda$^3$ \\
$^1$Indian Institute of Astrophysics, Bangalore 560034, India\\ 
$^2$Raman Research Institute, Bangalore 560080, India\\
$^3$Solar-Terrestrial Environment Laboratory, Nagoya University,Nagoya, Aichi, Japan}

\begin{document}
\date{Accepted--\hskip 1 cm   Received in original form --}

\pagerange{\pageref{firstpage}--\pageref{lastpage}} \pubyear{2008}

\maketitle

\label{firstpage}

\begin{abstract}
  The lower solar atmosphere consists of partially ionized turbulent plasmas  harbouring velocity field, magnetic field and current density fluctuations. The correlations amongst these small scale fluctuations give rise to large scale flows and magnetic fields which decisively affect all transport processes. The three fluid system consisting of electrons, ions and neutral particles supports nonideal effects such as the Hall effect and the ambipolar diffusion. Here, we study magnetic transport by ambipolar diffusion and compare the characteristic timescales of the laminar and the turbulent ambipolar diffusion 
processes. As expected from a turbulent transport process, the time scale of the turbulent ambipolar diffusion is found to be smaller by orders of magnitude as compared with the laminar ambipolar diffusion.The nonlinearity of the laminar ambipolar diffusion creates magnetic structures with sharp gradients which are amenable to processes such as magnetic reconnection and energy release therefrom for heating and flaring of the solar plasma.  
\end{abstract}

\maketitle

\section{Introduction}

The generation and the transport of magnetic fields, in astrophysical objects in general and in the sun in particular, is a subject of great interest and an active area of investigation. Several variants of the mean field dynamo have been suggested for the generation and migration of the magnetic flux \cite{ref1,ref2,ref3,ref4}. In a partially ionized plasma the collisions between the charged particles and the neutrals produce additional diffusion mechanisms such as the ambipolar diffusion due to ion-neutral collisions and the magnetic resistivity due to the electron-neutral collisions.

 The sun is endowed with a variety of temperature variations resulting from a combination of thermal and radiative equilibria and departures therefrom. The thermal and the nonthermal nature of processes then translates into a plasma with varying degrees of ionization. Thus the ionization fraction $\alpha=\frac{\rho_i}{\rho_n}$ could vary over several orders of magnitude where $\rho_i$ and $\rho_n$ are respectively the ion and the neutral hydrogen mass densities. Discrete structures such as sunspots, prominences and spicules contain plasmas with varying degrees of ionization. The support of the neutral component against gravity is a major concern in the stability of these structures. Although the ideal magnetohydrodynamics (MHD) is often used as a starting point of an investigation, a partially ionized system dominated by the charged particle-neutral collisions and the neutral particle dynamics necessitates a 3-fluid treatment.The strong charge particle-neutral coupling endows
  the neutral fluid with some of the properties of a conducting fluid. The neutral fluid is thus subjected to the Lorentz force along with the usual pressure gradient force. This attribute has been invoked to find the support for the neutral component of the  partially ionized cold and dense solar prominence\cite{ref5}. 
The evolution of the magnetic fields in such a
plasma would be affected by the multifluid interactions in general and the ambipolar diffusion in particular\cite{ref6}. The solar magnetic flux, generated in the convection zone, has to pass through the partially ionized solar  photosphere before it can appear high up in the solar corona. This realization is rather recent and is now receiving a lot of attention. Arber, Haynes and Leake\cite{ref7} has emphasized the profound effects on the temperature and the current structure of the overlying chromosphere and the corona that the inclusion of the neutral medium can produce. V\"{o}gler and Sch\"{u}ssler\cite{ref8} have invoked dynamo mechanism in a region a few hundred kilometers below and above the visible solar surface to account for the stronger horizontal magnetic fields. The issue of
possible disconnection between the sub-surface and the surface solar
magnetic field, recently emphasized by Sch\"ussler\cite{ref9}, may have some
bearing on the neglect of the neutral fluid-plasma coupling in the
flux transport on the solar photosphere. It is clear that the transport and or generation processes of the magnetic field on the solar photosphere must be studied in a partially ionized plasma. The magnetic transport would occur through the large scale flows as well as the turbulent fluctuations of the velocity field, the magnetic field and the current density with which the photosphere is well endowed. Recently Krishan and Gangadhara\cite{ref10} have initiated the study of mean-field dynamo in a partially ionized plasma.  

For a low degree of ionization one can define a weakly ionized plasma  by the condition\cite{ref11} that the electron-neutral collision frequency $\nu_{en}\sim 10^{-15}n_n(\frac{8K_BT}{\pi m_{en}})^{0.5}$  is much larger than the electron-ion collision frequency $\nu_{ei}\sim 6\times 10^{-24}n_i\Lambda Z^2(K_BT)^{-1.5}$. This translates into the ionization fraction $n_p/n_n < 5\times 10^{-11}T^2$ where $n's$ are the particle densities and $T$ is the temperature in Kelvin. A major part of the solar photosphere\cite{ref5} qualifies as a weakly ionized plasma\cite{ref13}.

In this paper, we investigate the magnetic transport by the flow generated by the magnetic field itself through the ambipolar diffusion in a self consistent manner. We develop a three fluid
framework in section two. The turbulent electromotive force and the large scale field dynamics is established in section 3. The advection of the large scale magnetic field, in the kinematic limit\cite{ref14} is studied and the timescales of advection by the laminar and the turbulent ambipolar diffusion are presented in section 4. We end the paper with a section on conclusion.

\section{Three-component magnetofluid}

We begin with the three component weakly ionized plasma consisting
of electrons (e), ions (i) of uniform mass density $\rho_i$ and
neutral particles (n) of uniform mass density $\rho_n$. The equation
of motion of the electrons can be written as:
\begin{eqnarray} 
&&m_{\rm e} n_{\rm e}\left[\frac{\partial {\b V}_{\rm e}}{\partial t}+ 
({\b V}_{\rm e}\cdot\nabla){\b V}_{\rm e}\right]=
 -\nabla p_{\rm e}- \nonumber \\
&&\quad\quad e n_{\rm e}\left[{\b E}+
 \frac{\b V_{\rm e}\times \b B}{c}\right]-
 m_{\rm e}n_{\rm e}\nu_{\rm en}(\b V_{\rm e}-\b V_{\rm n})~.
\end{eqnarray} 
where the electron-ion collisions have been neglected since the
ionized component is of low density.  On neglecting the electron
inertial force, the electric field $\b E$ is found to be:
\begin{eqnarray} 
{\b E}=-\frac{\b V_{\rm e}\times \b B}{c}-\frac{\nabla
    p_{\rm e}}{en_{\rm e}}-\frac{m_{\rm e}}{e}\nu_{\rm en}(\b V_{\rm e}-\b V_{\rm n})~.
\end{eqnarray}
This gives us Ohm's law. For time scales of interest larger than the ion-neutral collision time scale the ion
dynamics can be ignored. The ion force balance then becomes:
\begin{eqnarray}
0=-\nabla p_{\rm i}+e n_{\rm i}\left[{\b E}+\frac
{\b V_{\rm i}\times \b B}{c}\right]-\nu_{\rm in}\rho_{\rm i}(\b V_{\rm i}-\b V_{\rm n})~,
\end{eqnarray}
where $\nu_{\rm in} $ is the ion-neutral collision frequency, and the
ion-electron collisions have been neglected for the low density
ionized component.  Substituting for $\b E$ from Eq.~(2) we find the
relative velocity between the ions and the neutrals:
\begin{equation}
\b V_{\rm n}-\b V_{\rm i}= \frac{\nabla (p_{\rm i}+p_{\rm e})}{\nu_{\rm in} 
\rho_{\rm i}}-\frac{\b J\times \b B}{ c\nu_{\rm in}\rho_{\rm i}}, 
\end{equation}
where
\begin{equation}
\b J = en_{\rm e}(\b V_{\rm i}-\b V_{\rm e})~.
\end{equation}
  The equation of motion of the neutral fluid is:
\begin{eqnarray}
  \rho_{\rm n} \left[ \frac{\partial \b V_{\rm n}}{\partial t}+({\b V_{\rm n}} \cdot\nabla )
{\b V_{\rm n}}\right]&=& -\nabla p_{\rm n}-\nu_{\rm ni}\rho_{\rm n}(\b V_{\rm n}-\b V_{\rm i})-\nonumber\\
&& \nu_{\rm ne}\rho_{\rm n}(\b V_{\rm n}-\b V_{\rm e})~, 
\end{eqnarray}
where the viscosity of the neutral fluid has been neglected.
Substituting for $\bm V_{\rm n}-\bm V_{\rm i}$ from Eq. (4), and using $\nu_{\rm in}
\rho_{\rm i}= \nu_{\rm ni} \rho_{\rm n}$ we find:
\begin{eqnarray}
\rho_{\rm n} \left[ \frac{\partial \b V_{\rm n}}{\partial t}+(\b V_{\rm n} 
\cdot\nabla)\b V_{\rm n}\right]=-\nabla p+\frac{{\b J}\times {\b B}}{c}~,
\end{eqnarray}
where $p=p_{\rm n}+p_{\rm i}+p_{\rm e}$.  
Observe that the neutral fluid is subjected to the Lorentz force as a
result of the strong ion-neutral coupling due to their collisions. 

Consider Faraday's law of induction:
\begin{equation}
\frac{\partial \b B}{\partial t}=-c\nabla\times\b E
\end{equation}
By substituting for the electric field from Eq.~(2), we get
\begin{equation}
\frac{\partial \b B}{\partial t}=\nabla\times(\b V_{\rm e}\times\b B)+
\eta \nabla^2 \b B~,
\end{equation}
where the pressure gradient terms have been dropped for the incompressible
case with constant temperature.
Here $\eta=m_{\rm e}\nu_{\rm en}c^2/(4\pi e^2n_{\rm e})$ is the electrical
resistivity predominantly due to electron-neutral collisions.
Using the construction
\begin{equation}
\b V_{\rm e}\times \b B=[\b V_{\rm n}-(\b V_{\rm n}-\b V_{\rm i})-
                        (\b V_{\rm i}-\b V_{\rm e})]\times\b B~,
\end{equation}
and substituting for the relative velocity of the ion and the neutral fluid
from Eq.~(4), Eq.~(9) becomes:
\begin{eqnarray}
{\partial \b B\over\partial t}=\nabla\times\left[\left(\b V_{\rm n} 
-\frac{\b J}{en_{\rm e}}+\frac{\b J\times\b B}{ c\nu_{\rm in}\rho_{\rm i}}
\right)\times \b{B}\right]+\eta{\nabla}^{2}\b B
\end{eqnarray}
One can easily identify the Hall term ($\b J/e n_{\rm e}$), and the
ambipolar diffusion term ($\b J\times \b B $)\cite{ref7} 
. The Hall term is much larger than the ambipolar term for large
neutral particle densities or for $\nu_{\rm in} \gg \omega_{\rm ci}$
where $\omega_{\rm ci}$ is the ion cyclotron frequency. In this system
the magnetic field is not frozen to any of the fluids. Equations (7)
and (11) along with the mass conservation
\begin{eqnarray}
\nabla\cdot\b V_{\rm n}=0
\end{eqnarray}
form the basis of our investigation. 

\section{Turbulence in three-component magnetofluid}
One way of studying turbulence in a system is by splitting the physical quantities into small scale and large scale parts. The mean-field dynamo\cite{ref4} is a consequence of such a study. We will tread the same path.  The magnetic induction equation (11) is 
written as:
\begin{eqnarray}
{\partial \b B\over\partial t}=\nabla\times\left[\b V_E 
\times \b{B}\right]+\eta{\nabla}^{2}\b B~,
\end{eqnarray}
where
\begin{equation}
\b V_{\rm E}= \b V_{\rm n}+\b V_{\rm H}+\b V_{\rm Am}
\end{equation}
with
\begin{eqnarray}
\b V_{\rm H}= -\frac{\b J}{en_{\rm e}}
\end{eqnarray}
as the Hall velocity and 
\begin{eqnarray}
\b V_{\rm Am}= \frac{\b J\times\b B}{ c\nu_{\rm in}\rho_{\rm i}}
\end{eqnarray}
could be called the ambipolar velocity. Following the standard
procedure\cite{ref4} the velocity $\b V_{\rm E}$ and the
magnetic field $\b B$ are split into their average large scale parts
and the fluctuating small scale parts as:
\begin{eqnarray}
\b V_{\rm E} &=& \oline{\b V_{\rm E}}+\b V'_{\rm E}, \\  
       \b B &=& \oline{\b B}+\b B'
\end{eqnarray}
such that
\begin{eqnarray}
\oline{\b V'_{\rm E}}=0, \quad\quad \oline{\b B'}=0.
\end{eqnarray} 
In the kinematic dynamo the magnetic induction equation is solved for
large and small scale fields.  Substituting Eqs.~(17) and (18) into
the induction equation (11), we find, in the first order smoothing
approximation,
\begin{eqnarray}
\b V'_{\rm E}=\b V'_n-\frac{\b J'}{en_{\rm e}}+\frac{\b J'\times\oline {\b B}}
{c\nu_{\rm in}\rho_{\rm i}}
+ \frac{\oline{\b J}\times\b B'}{c\nu_{\rm in}\rho_{\rm i}}
\end{eqnarray}
and the mean flow is found to be:
\begin{eqnarray}
\oline{\b V_{\rm E}}=\oline{\b V_{\rm n}}-\frac{\oline {\b J}}{en_{\rm e}}+
             \frac{\oline{\b J}\times\oline{\b B}}{c\nu_{\rm in}\rho_{\rm i}}+\frac{\oline{\b J'\times\b B'}}{c\nu_{\rm in}\rho_{\rm i}}~.
\end{eqnarray}
The turbulent electromotive force $\Xi$ is a function of the mean
magnetic induction $\oline{\b B}$ and mean quantities formed from the
fluctuations, and is expressed as:
\begin{eqnarray}
\Xi =\oline{\b V'_{\rm E}\times\b B'} = \alpha \oline{\b B} - 
     \beta\, \nabla\times\oline{\b B}~,
\end{eqnarray}
where
\begin{eqnarray}
\alpha &=& - \frac{\tau_{\rm cor}}{3}\oline{\b V'_{\rm E}\cdot
                  (\nabla\times\b V'_{\rm E})}\nonumber\\
       &=& \alpha_{\rm v}+\alpha_{\rm H}+\alpha_{\rm Am}~.
\end{eqnarray}
Here
\begin{eqnarray}
\alpha_{\rm v} = -\frac{\tau_{\rm cor}}{3}\oline{\b V'_{\rm n} \cdot\Omega'_{\rm n}}
\end{eqnarray}
is the measure of the average kinetic helicity of the neutral fluid in
the turbulence possessing correlations over time $\tau_{cor}$ and
\begin{eqnarray}
\alpha_{\rm H} = \frac{2\tau_{\rm cor}}{3en_{\rm e}}\oline{\b J'
                 \cdot\b\Omega'_{\rm n}}
\end{eqnarray}
represents the contribution of the Hall effect. The coupling of the
charged components with the neutral fluid is clearly manifest through
the possible correlation between the current density fluctuations and
the vorticity fluctuations of the neutral fluid
$\b \Omega'_{\rm n}=\nabla\times \b V'_{\rm n} .$ The ambipolar term gives 
rise to  
\begin{equation}
\alpha_{\rm Am} = \b \alpha_{\rm A}\cdot\oline{\b B}~,
\end{equation}
with
\begin{equation}
\b \alpha_{\rm A} = \frac{2\tau_{\rm cor}}{3c\rho_{\rm i}\nu_{\rm in}}\oline
                    {\b J'\times\b\Omega'_{\rm n}}~,
\end{equation}
as the contribution from the ambipolar diffusion with its essential
nonlinear character manifest through its dependence on the average
magnetic induction. One also observes that the Hall alpha (Eq. 25)
requires a component of the fluctuating current density along the
fluctuating vorticity of the neutral fluid whereas the ambipolar
effect (Eq. 27) thrives on the component of the fluctuating current
density perpendicular to the fluctuating vorticity. The turbulent
dissipation is given by
\begin{eqnarray}
\beta = \frac{\tau_{\rm cor}}{3}\oline{{\b V}'^2_{\rm E}}
= \beta_{\rm v}+\beta_{\rm H}+\beta_{\rm Am}
\end{eqnarray}
with
\begin{eqnarray}
\beta_{\rm v} = \frac{\tau_{\rm cor}}{3}\oline{{\b V}'^2_{\rm n}}
\end{eqnarray}
as the measure of the average turbulent kinetic energy of the neutral
fluid in the turbulence possessing correlations over time
$\tau_{\rm cor}$ and
\begin{eqnarray}
\beta_{\rm H} =- \frac{2\tau_{\rm cor}}{3en_{\rm e}}\oline{\b J' \cdot\b V'_{\rm n}}
\end{eqnarray}
represents the contribution of the Hall effect. The coupling of the
charged components with the neutral fluid is clearly manifest through
the possible correlation between the current density fluctuations and
the velocity fluctuations of the neutral fluid. The ambipolar term furnishes
\begin{equation}
\beta_{\rm Am} = \b \beta_A\cdot \oline{\b B} ~,
\end{equation}
\begin{equation}
\b \beta_{\rm A} = -\frac{2\tau_{\rm cor}}{3c\rho_i\nu_{\rm in}}\oline{\b 
                    J'\times\b V'_{\rm n}}
\end{equation}
with its essential
nonlinear character manifest through its dependence on the average
magnetic induction. One also observes that the Hall $\beta_{\rm H}$ requires
a component of the current density fluctuations along the velocity
fluctuations of the neutral fluid whereas the ambipolar effect thrives
on the component of the current density fluctuations perpendicular to
the velocity fluctuations. We have used rigid or perfectly conducting
boundary conditions (all surface contributions vanish) while
determining the averages. The magnetic transport equation becomes:
\begin{equation}
{\partial\oline{ \b B}\over\partial t}=\nabla\times\left[\oline{\b V_{\rm E}}\times\oline{\b B}+\alpha{\oline{\b B}}
-\beta \nabla\times {\oline{\b B}}\right]+ \eta{\nabla}^{2}{\oline{\b B}}~.
\end{equation}

\section{Magnetic transport by turbulent ambipolar diffusion}
 In order to highlight the effect of the ambipolar diffusion on the magnetic transport, We ignore all other effects so that the large scale neutral fluid flow $\oline{\b V_{\rm n}}=0$, the Hall flow $\oline{\b V_{\rm H}}=-\frac{\oline {\b J}}{en_{\rm e}}=0$, $\alpha=0$, $\beta=0$ and $\eta=0$. The transport equation reduces to

\begin{equation}
{\partial\oline{ \b B}\over\partial t}=\nabla\times\left[\frac{\oline{\b J}\times\oline{\b B}}{c\nu_{\rm in}\rho_{\rm i}}\times\oline{\b B}+\frac{\oline{\b J'\times\b B'}}{c\nu_{\rm in}\rho_{\rm i}}\times\oline{\b B}\right]~.
\end{equation}
 The first term on the right hand side represents the laminar ambipolar diffusion. The second term represents the turbulent ambipolar diffusion or transport of the magnetic induction. We estimate the second term as follows:
\begin{equation}
\oline{\b J'\times\b B'}=\oline{\frac{-c\nabla {B'}^2}{8\pi }+c\frac{\b B'.\nabla \b B'}{4\pi }}~.
\end{equation}
The left hand side is the average Lorentz force due to magnetic fluctuations and we assume that it is nonzero. On the right side, the first term is symptomatic of the turbulent magnetic pressure and the second is the contribution to the turbulent curvature. If  magnetic fluctuations are introduced in a fluid initially at rest, the Lorentz force will drive motion in the fluid. In an incompressible fluid the direction of the flow is along the predominant component of the fluctuating field. This arises through the curvature term $\b B'.\nabla\b B'$ since only the solenoidal part of the Lorentz force can drive flows in an incompressible fluid or in a compressible fluid under the Boussinesq approximation as discussed by Ogilvie\cite{ref14}. The gradient part of the Lorentz force is compensated by the pressure gradient. Here we shall make an approximate estimate of this term as:
\begin{equation}
\oline{\b J'\times\b B'}\approx\frac{c}{4\pi\lambda_{cor}}\oline{ B'^{2}}~.
\end{equation}
 where $\lambda_{cor}$ is the correlation length of the magnetic fluctuations. We follow \cite{ref4} in order to estimate the mean square of the magnetic fluctuations $\overline{ B'^{2}}$. In the high conductivity limit ($\eta\approx 0$), the fluctuation $\b B'$ is produced by the interaction of the turbulent velocity field $\b V'_{E}$ with the mean magnetic field $\oline{ \b B}$ over a time scale $\tau_{cor}$ where $\tau_{cor}$ is the correlation timescale of the fluctuating fields. Thus the elemental field $B'_{el}$ is given by 
\begin{equation}
B'_{el}\approx \frac{V'_E\tau_{cor}}{\lambda_{cor}}\oline B~.
\end{equation}
After a time interval $\tau_{cor}$ the turbulent field $\b V'_E$ changes nearly completely. This new realization of the turbulent $\b V'_E$ interacts with $\oline{\b B}$ to produce another elemental  field $B'_{el}$ uncorrelated with the previous field element. Each field element survives over the dissipation time scale $ \tau_{dis}\approx\frac{\lambda^2_{cor}}{\eta}$ much larger than the correlation time $\tau_{cor}$. Thus the total fluctuating field $B'$ is the incoherent sum of these elemental fields $B'_{el}$. The number $n$ of these elemental fields is given by
\begin{equation}
n\approx\frac{\lambda^2_{cor}}{\eta\tau_{cor}}~.
\end{equation}
Thus, to the order of magnitude, $B'$ is found to be
\begin{equation}
B'\approx \sqrt{n} B'_{el}\approx\sqrt{\frac{{V'_E}^2\tau_{cor}}{\eta}}\oline B~.
\end{equation}
And
\begin{equation}
\oline{{B'}^2}\approx R_M S{\oline B}^2~.
\end{equation}
where 
\begin{equation}
R_M=\frac{V'_E\lambda_{cor}}{\eta},\\ S=\frac{V'_E\tau_{cor}}{\lambda_{cor}}~.
\end{equation}
are respectively the magnetic Reynolds number and the Strouhal number. We can now estimate the timescales of the laminar ambipolar diffusion, $T_{LA}$  and the turbulent ambipolar diffusion, $T_{TA}$ from the magnetic transport equation (Eq.34) as: 
\begin{equation}
T_{LA}\approx \frac{4\pi L^2\rho_i\nu_{in}}{{\oline B}^2},\\T_{TA}\approx \frac{4\pi L\lambda_{cor}\rho_i\nu_{in}}{{\oline {{B'}^2}}}~.
\end{equation}
where $L$ is the characteristic spatial scale associated with the mean magnetic field $\oline B$.
Substituting from Eq.(40), we find the ratio of the two timescales to be:
\begin{equation}
\frac{T_{TA}}{T_{LA}}=(R_M S)^{-1}\frac{\lambda_{cor}}{L}~.
\end{equation}
Now the magnetic Reynolds number $R_M>>1$, the Strouhal number $S<1$ and the correlation length $\lambda_{cor}<<L$; therefore the turbulent ambipolar diffusion timescale $T_{TA}$ can be much smaller than the laminar ambipolar diffusion timescale $T_{LA}$. This is exactly what is expected of a turbulent transport process. The reason for the greater efficiency of a turbulent transport over a laminar transport is that in a turbulent process a physical attribute on a large scale is transferred to a small scale and it is much easier for a small scale quantity to diffuse than for a large scale one. For the typical values of the physical parameters at a height of $500 Km$ on the solar atmosphere we find $T_{LA}\approx 10^6 sec$ and $T_{TA}$ could be smaller by several orders of magnitude. 
\section{Conclusion}
It is found that the ambipolar diffusion is an important effect in the partially ionized part of the solar atmosphere. The laminar ambipolar transport is known to create magnetic structures \cite{ref6} with steep gradients\cite{15}. The turbulent part of the ambipolar diffusion could transport these structures on a timescale  which could be shorter by orders of magnitude than all other processes. Thus the formation and migration of magnetic structures to higher up in the solar atmosphere could be the result of ambipolar diffusion.

\section*{Acknowledgments}The authors are grateful to Dr. B.A. Varghese for his help in the preparation of this manuscript.

\end{document}